\documentclass[twocolumn,showpacs,preprintnumbers,superscriptaddress]{revtex4}
\usepackage{graphicx,bm,times}
\graphicspath{{./fig/}{./png/}}
\newcommand{\BoldVec}[1]{\mathchoice%
  {\mbox{\boldmath $\displaystyle     #1$}}%
  {\mbox{\boldmath $\textstyle        #1$}}%
  {\mbox{\boldmath $\scriptstyle      #1$}}%
  {\mbox{\boldmath $\scriptscriptstyle#1$}}%
}
\newcommand{\EQ}{\begin{equation}}
\newcommand{\EN}{\end{equation}}
\newcommand{\EQA}{\begin{eqnarray}}
\newcommand{\ENA}{\end{eqnarray}}
\newcommand{\eq}[1]{(\ref{#1})}

\newcommand{\Eq}[1]{Eq.~(\ref{#1})}

\newcommand{\eqs}[2]{(\ref{#1}) and~(\ref{#2})}

\newcommand{\Fig}[1]{Fig.~\ref{#1}}

\newcommand{\Tab}[1]{Table~\ref{#1}}

\newcommand{\bra}[1]{\langle #1\rangle}

\newcommand{\meanB}{\overline{B}}

\newcommand{\meanE}{\overline{E}}
\newcommand{\meanJ}{\overline{J}}

\newcommand{\meanPhi}{\overline{\Phi}}

\newcommand{\meanAA}{\overline{\mbox{\boldmath $A$}}}
\newcommand{\meanBB}{\overline{\bm{B}}}
\newcommand{\meanEE}{\overline{\bm{E}}}
\newcommand{\meanUU}{\overline{\bm{U}}}

\newcommand{\meanJJ}{\overline{\mbox{\boldmath $J$}}}

\newcommand{\meanemf}{\overline{\cal E} {}}

\newcommand{\meanEMF}{\overline{\mbox{\boldmath ${\cal E}$}} {}}

{}
{}
\newcommand{\meanFFFF}{\overline{\mbox{\boldmath ${\cal F}$}}{}}{}

\newcommand{\epsm}{\epsilon_{\rm m} {}}

\newcommand{\nullvector}{{\bf0}}
\newcommand{\nnn}{\hat{\mbox{\boldmath $n$}} {}}

\newcommand{\uu}{\BoldVec{u} {}}

\newcommand{\UU}{\BoldVec{U} {}}
\newcommand{\bb}{\BoldVec{b} {}}

\newcommand{\BB}{\BoldVec{B} {}}

\newcommand{\AAA}{\BoldVec{A} {}}
\newcommand{\aaa}{\BoldVec{a} {}}
\newcommand{\aaaa}{\BoldVec{a} {}} 

\newcommand{\jj}{\BoldVec{j} {}}
\newcommand{\JJ}{\BoldVec{J} {}}

\newcommand{\EE}{\BoldVec{E} {}}

\newcommand{\nab}{\BoldVec{\nabla} {}}

\newcommand{\dd}{{\rm d} {}}
\newcommand{\const}{{\rm const}  {}}

\def\Rm{R_{\rm m}}
\def\vA{v_{\rm A}}

\def\RRey{{\cal R}}

\def\LLu{{\cal L}}

\def\kf{k_{\rm f}}
\def\kmean{k_{\rm m}}
\def\urms{u_{\rm rms}}

\def\etat{\eta_{\rm t}}
\def\etaT{\eta_{\rm T}}
\def\Beq{B_{\rm eq}}

\def\half{{\textstyle{1\over2}}}

\def\onethird{{\textstyle{1\over3}}}

\newcommand{\yan}[3]{, Astron. Nachr. {\bf #2}, #3 (#1).}

\newcommand{\yana}[3]{, Astron. Astrophys. {\bf #2}, #3 (#1).}

\newcommand{\ymn}[3]{, Mon.\ Not.\ R.\ Astron.\ Soc.\ {\bf #2}, #3 (#1).}

\newcommand{\yjfm}[3]{, J. Fluid Mech. {\bf #2}, #3 (#1).}

\newcommand{\yprl}[3]{, Phys.\ Rev.\ Lett.\ {\bf #2}, #3 (#1).}

\newcommand{\yjgr}[3]{, J. Geophys. Res. {\bf #2}, #3 (#1).}

\newcommand{\yapj}[3]{, Astrophys. J. {\bf #2}, #3 (#1).}

\newcommand{\ypp}[3]{, Phys. Plasmas {\bf #2}, #3 (#1).}
\newcommand{\yppcf}[3]{, Plasmas Phys. Contr. Fusion {\bf #2}, #3 (#1).}

\newcommand{\ypf}[3]{, Phys. Fluids {\bf #2}, #3 (#1).}

\newcommand{\ygafd}[3]{, Geophys. Astrophys. Fluid Dynam. {\bf #2}, #3 (#1).}

\newcommand{\yjour}[4]{, #2 {\bf #3}, #4 (#1).}

\newcommand{\ybook}[3]{, {\em #2}. #3 (#1).}

\begin{document}
\title{A model of driven and decaying magnetic turbulence in a cylinder}
\author{Koen Kemel}
\affiliation{NORDITA, AlbaNova University Center, Roslagstullsbacken 23,
SE-10691 Stockholm, Sweden}
\affiliation{Department of Astronomy,
Stockholm University, SE 10691 Stockholm, Sweden}

\author{Axel Brandenburg}
\affiliation{NORDITA, AlbaNova University Center, Roslagstullsbacken 23,
SE-10691 Stockholm, Sweden}
\affiliation{Department of Astronomy,
Stockholm University, SE 10691 Stockholm, Sweden}

\author{Hantao Ji}
\affiliation{Center for Magnetic Self-Organization in Laboratory and
Astrophysical Plasmas, Princeton Plasma Physics Laboratory, Princeton
University, Princeton, New Jersey 08543, USA}

\date{\today,~ $ $Revision: 1.59 $ $}
\begin{abstract}
Using mean-field theory, we compute the evolution of the magnetic field
in a cylinder with outer perfectly conducting boundaries, an imposed
axial magnetic and electric field.
The thus injected magnetic helicity in the system can be redistributed
by magnetic helicity fluxes down the gradient of the local current helicity
of the small-scale magnetic field.
A weak reversal of the axial magnetic field is found to be a consequence
of the magnetic helicity flux in the system.
Such fluxes are known to alleviate so-called catastrophic quenching of
the $\alpha$ effect in astrophysical applications.
Application to the reversed field pinch in plasma confinement
devices is discussed.
\end{abstract}

\maketitle

\section{Introduction}

The interaction between a conducting medium moving at speed $\UU$ through
a magnetic field $\BB$ is generally referred to as a dynamo effect.
This effect plays important roles in astrophysics \citep{Mof78,KR80},
magnetospheric physics \citep{Ogino},
as well as laboratory plasma physics \citep{Ji_etal96}.
It modifies the electric field in the rest frame, so that Ohm's law
takes the form $\JJ=\sigma\left(\EE+\UU\times\BB\right)$,
where $\JJ$ is the current density, $\EE$ is the electric field,
and $\sigma$ is the conductivity.
Of particular interest for the present paper is the case where an external
electric field $\EE^{\rm ext}$ is induced through a transformer with a
time-varying magnetic field, as is the case in many plasma confinement
experiments.
With the external electric field included, Ohm's law becomes
\EQ
\JJ=\sigma\left(\EE+\EE^{\rm ext}+\UU\times\BB\right).
\EN
In a turbulent medium,
often only averaged quantities (indicated below by overbars) are accessible.
The averaged form of Ohm's law reads
\EQ
\meanJJ=\sigma\left(\meanEE+\meanEE^{\rm ext}+\meanUU\times\meanBB
+\meanEMF\right),
\label{OhmMean}
\EN
where $\meanEMF=\overline{\uu\times\bb}$ is referred to as the mean
$\meanEMF=\overline{\uu\times\bb}$ is referred to as the mean
turbulent electromotive force, and $\uu=\UU-\meanUU$ and $\bb=\BB-\meanBB$
are fluctuations of velocity and magnetic field, respectively.
It has been known for some time that the averaged profiles,
$\meanJJ$ and $\sigma\meanEE^{\rm ext}$ do not agree in actual experiments.
This disagreement cannot be explained by the $\meanUU\times\meanBB$
term either, leaving therefore $\meanEMF$ as the only remaining term.
Examples include the recent dynamo experiment in Cadarache
\cite{Cadarache} and in particular the reversed field pinch (RFP)
\cite{BN80,Tay86,Ji_etal96}, which is one of the configurations studied
in connection with fusion plasmas.
The name of this device derives from the fact that the toroidal
(or axial, in a cylindrical geometry) magnetic field
reverses sign near the periphery.
Indeed, in the astrophysical context it is well-known that the
$\meanEMF$ is responsible for the amplification and maintenance of
large-scale magnetic fields \citep{Mof78,KR80}.

The analogy among the various examples of the $\meanEMF$ term
has motivated comparative research between astrophysics and plasma
physics applications \cite{BJ06}.
In these cases, $\meanEMF$ is found to have a component
proportional to the mean field ($\alpha\meanBB$, referred
to as the $\alpha$ effect) and a component proportional to the
mean current density ($\etat\meanJJ$, where $\etat$ is the turbulent diffusivity).
Since $\alpha$ is a pseudoscalar, one expects it to depend on the
helicity of the flow, which is also a pseudoscalar.
Decisive in developing the analogy between the $\alpha$ effects in
astrophysics and laboratory plasma physics is the realization that $\alpha$
is caused not only by helicity in the flow (kinetic $\alpha$ effect),
but also by that of the magnetic field itself \cite{PFL76}.
This magnetic contribution to the $\alpha$ effect has received increased
astrophysical interest, because there are strong indications that
such dynamos saturate by building up small-scale helical fields
that lead to a magnetic $\alpha$ effect which, in turn,
counteracts the kinetic $\alpha$ effect \cite{FB02,BB02,Sub02}.
This process can be described quantitatively by taking
magnetic helicity evolution into
account, which leads to what is known as the dynamical quenching formalism
that goes back to early work of Kleeorin \& Ruzmaikin \cite{KR82}.
However, it is now also believed that such quenching would lead to
a catastrophically low saturation field strength \cite{BS05}, unless there are
magnetic helicity fluxes out of the domain that would
limit the excessive build-up of small-scale helical fields \cite{BS04}.
This would reduce the magnetic $\alpha$ effect and thus
allow the production of mean fields whose energy density is comparable
to that of the kinetic energy of the turbulence \cite{B05}.

These recent developments are purely theoretical, so the hope
is that more can be learnt by applying the recently gained knowledge
to experiments like the RFP \cite{BN80,Tay86}.
Unlike tokamaks, the RFP is a relatively slender torus, so it makes
sense to study its properties in a local model where one ignores
curvature effects and considers a cylindrical piece of the torus.
Along the axis of this cylinder there is
a field-aligned current that makes the field helical.
This field is susceptible to kink and tearing instabilities
that lead to turbulence.
It is generally believed that the resulting mean turbulent electromotive
force $\meanEMF$ is responsible for the field reversal
\cite{Ji_etal96,JP02}.
The turbulence is also believed to help driving the system
toward a minimum energy state \cite{Tay74}.
This state is nearly force-free and maintained by $\meanEE^{\rm ext}$.
This adds to the notion that the RFP must be sustained by some kind
of dynamo process \cite{AB84}.
In Cartesian geometry such a slow-down has previously already been
modeled using the dynamical quenching formalism \cite{YBR03}.

The RFP has been studied extensively using three-dimensional simulations
\cite{AB84,SCN85,SMN96,CBE06}, which confirm the conjecture of J.\ B.\
Taylor \cite{Tay74} that the system approaches a minimum energy state.
Additional understanding has been obtained using mean-field considerations
\cite{Str85,BH86}.
Both, here and in astrophysical dynamos
there is an $\alpha$ effect that quantifies the
correlation of the fluctuating parts of velocity and magnetic field.
However, a major difference lies in the fact that in the RFP the
$\alpha$ effect is caused by instabilities of the
initially large-scale magnetic field while
in the astrophysical case one is concerned with the problem of explaining
the origin of large-scale fields by the $\alpha$ effect \cite{Mof78,KR80}.
However, this distinction may be too simplistic and there is indeed
evidence that in the RFP the $\alpha$ effect exists in close relation
with a finite magnetic helicity flux \cite{Ji99}, supporting the idea
that so-called catastrophic quenching is avoided by helicity transport.

The purpose of this paper is to apply modern mean-field dynamo theory
with dynamical quenching to a cylindrical configuration to allow a more
meaningful comparison between the $\alpha$ effect in astrophysics and
the one occurring in RFP experiments.

\section{The model}

To model the evolution of the magnetic field in a cylinder with
imposed axial magnetic and electric fields, we employ mean-field theory,
where the evolution of the mean field $\meanBB$ is governed by turbulent
magnetic diffusivity and an $\alpha$ effect.
Unlike the astrophysical case where $\alpha$ depends primarily on the
kinetic helicity of the plasma, in turbulence from current-driven
instabilities the $\alpha$ effect is likely to depend
primarily on the current helicity of the small-scale field \cite{PFL76}.
The current density is given by $\JJ=\nab\times\BB/\mu_0$, where
$\mu_0$ is the vacuum permeability and $\jj=\nab\times\bb/\mu_0$ is the
fluctuating current density.
The mean current helicity density of the small-scale field is then given by
$\overline{\jj\cdot\bb}$.
To a good approximation, the $\overline{\jj\cdot\bb}$ term is proportional
to the small-scale magnetic helicity density, $\overline{\aaaa\cdot\bb}$,
where $\aaaa=\AAA-\meanAA$ is the vector potential of the fluctuating field.
The generation of $\overline{\aaaa\cdot\bb}$ is coupled to the decay of
$\meanAA\cdot\meanBB$ through the magnetic helicity evolution equation
\cite{KR82,KRR95,FB02,BB02} such that
$\meanAA\cdot\meanBB+\overline{\aaaa\cdot\bb}$ evolves only resistively
in the absence of magnetic helicity fluxes.

Note that $\overline{\aaa\cdot\bb}$ is in general gauge-dependent and
might therefore not be a physically meaningful quantity.
However, if there is sufficient scale separation, the mean magnetic
helicity density of the fluctuating field can be expressed in terms of
the density of field line linkages, which does not involve the magnetic
vector potential and is therefore gauge-independent \cite{SB06}.
For the large-scale field, on the other hand,
the magnetic helicity density does remain in general
gauge-dependent \cite{HB10}, but our final model will not be affected
by this, because the magnetic helicity of the large-scale magnetic field
does not enter in the mean-field model.

We model an induced electric field by an externally applied
electric field $\EE^{\rm ext}$.
In the absence of any other induction effects this leads to a current
density $\JJ=\sigma\EE^{\rm ext}$.
Furthermore, we ignore a mean flow ($\meanUU=\nullvector$), and assume
that the velocity field has only a turbulent component $\uu$.
For simplicity we assume that $\meanEE^{\rm ext}$ has no fluctuating part,
i.e.\ $\EE^{\rm ext}=\meanEE^{\rm ext}$.
The decay of $\meanBB$ is accelerated by turbulent magnetic
diffusivity $\etat$, which is expected to occur as a result of the
turbulence connected with kink and tearing instabilities inherent to the RFP.
This mean turbulent electromotive force has two components corresponding
to the $\alpha$ effect and turbulent diffusion with
\EQ
\meanEMF=\alpha\meanBB-\etat\mu_0\meanJJ,
\label{meanEMF}
\EN
where we have ignored the fact that $\alpha$ effect and turbulent
diffusion are really tensors.
The evolution equation for $\meanBB$ is then given by the mean-field
induction equation,
\EQ
{\partial\meanBB\over\partial t}=
\nab\times\left(\alpha\meanBB-\etaT\mu_0\meanJJ+\meanEE^{\rm ext}\right),
\label{dmeanBdt}
\EN
where $\etaT=\etat+\eta$ is the sum of turbulent and microscopic
(Spitzer) magnetic diffusivities (not to be confused with the resistivity
$\eta\mu_0$, which is also often called $\eta$).
Note that only non-uniform and non-potential contributions to
$\meanEE^{\rm ext}$ can have an effect.

As a starting point, we assume that the rms velocity $\urms$ and the
typical wavenumber $\kf$ of the turbulence are constant, although it is
clear that these values should really depend on the level of the actual
magnetic field.
We estimate the value of $\etat$ using a standard formula for isotropic
turbulence,
\EQ
\etat=\onethird\tau\overline{\uu^2},
\EN
where $\tau=(\urms\kf)^{-1}$ is the correlation time of the turbulence
and $\urms=(\overline{\uu^2})^{1/2}$ is its rms velocity.
Thus, we can also write $\etat=\urms/3\kf$.
The turbulent velocity results from kink and tearing mode instabilities
and will simply be treated as a constant in our model.
For the $\alpha$ effect we assume that the kinetic helicity is negligible
It would be much smaller than the current helicity, but of the same sign
\cite{PFL76}, so it would contribute to quenching the $\alpha$ effect,
and so we just take
\EQ
\alpha=\onethird\tau\overline{\jj\cdot\bb}/\rho_0,
\EN
and use the fact that $\overline{\jj\cdot\bb}$ and $\overline{\aaaa\cdot\bb}$
are proportional to each other.
Here, $\rho_0$ is the mean density of the plasma.
For homogeneous turbulence we have
$\overline{\jj\cdot\bb}=\kf^2\overline{\aaaa\cdot\bb}/\mu_0$,
although for inhomogeneous turbulence,
$\kf^2\overline{\aaaa\cdot\bb}/\mu_0$ has been found to be smaller than
$\overline{\jj\cdot\bb}$ by a factor of two \cite{Mit10}.
We compute the evolution of $\overline{\aaaa\cdot\bb}$ by considering
first the evolution equation for $\overline{\AAA\cdot\BB}$.
Note that $\overline{\AAA\cdot\BB}$ evolves only resistively, unless there
is material motion through the domain boundaries \cite{HB10}, so we have
\EQ
{\dd\over\dd t}\overline{\AAA\cdot\BB}=
2\meanEE^{\rm ext}\cdot\meanBB-2\eta\mu_0\overline{\JJ\cdot\BB}
-\nab\cdot\meanFFFF,
\label{dmeanHdt}
\EN
where $\meanFFFF$ is the mean magnetic helicity flux.
While $\meanBB$ evolves subject to the mean field equation \eq{dmeanBdt},
the magnetic helicity of the mean field will change subject to the equation
\EQ
{\dd\over\dd t}\left(\meanAA\cdot\meanBB\right)=
2\meanEMF^{\rm tot}\cdot\meanBB-2\eta\mu_0\meanJJ\cdot\meanBB
-\nab\cdot\meanFFFF_{\rm m},
\label{dAABBdt}
\EN
where $\meanEMF^{\rm tot}=\meanEMF+\meanEE^{\rm ext}$ and
$\meanFFFF_{\rm m}=\meanEE\times\meanAA+\meanPhi\,\meanBB$
is the mean magnetic helicity flux from the mean magnetic field,
and $\meanPhi$ is the mean electrostatic potential.
Here, $\meanEE=\eta\mu_0\meanJJ-\meanEMF^{\rm tot}$ is the mean electric field.
Subtracting \eq{dAABBdt} from \eq{dmeanHdt},
we find a similar evolution equation for $\overline{\aaaa\cdot\bb}$,
\EQ
{\dd\over\dd t}\overline{\aaaa\cdot\bb}=
-2\meanEMF\cdot\meanBB-2\eta\mu_0\overline{\jj\cdot\bb}
-\nab\cdot\meanFFFF_{\rm f},
\label{dabdt}
\EN
where $\meanFFFF_{\rm f}=\meanFFFF-\meanFFFF_{\rm m}$
is the mean magnetic helicity flux from the fluctuating magnetic field.
Note that $\EE^{\rm ext}$ does not enter in \Eq{dabdt}, because
$\EE^{\rm ext}=\meanEE^{\rm ext}$ has no fluctuations.
This equation can readily be formulated as an evolution equation for $\alpha$
by writing $\alpha=(\tau\kf^2/3\rho_0\mu_0)\overline{\aaaa\cdot\bb}$, i.e.,
\EQ
{\partial\alpha\over\partial t}=-2\etat\kf^2\meanEMF\cdot\meanBB/B_{\rm eq}^2
-2\eta\kf^2\alpha-\nab\cdot\meanFFFF_\alpha,
\label{dalpdt}
\EN
where $\meanFFFF_\alpha=(\tau\kf^2/3\rho_0\mu_0)\meanFFFF_{\rm f}$,
which is a rescaled magnetic helicity flux of the small-scale field
and $B_{\rm eq}$ is the field strength for which magnetic and
kinetic energy densities are equal, i.e.,
\EQ
B_{\rm eq}^2=\mu_0\rho_0\urms^2=(3\rho_0\mu_0/\tau)\,\etat.
\EN
We recall that
in the astrophysical context, equation \eq{dalpdt} is referred to as
the dynamical quenching model \cite{KR82,KRR95}.
In a first set of models we assume $\meanFFFF_\alpha=\nullvector$,
but later we shall allow for the fluxes to obey a Fickian diffusion law,
\EQ
\meanFFFF_\alpha=-\kappa_\alpha\nab\alpha,
\EN
where $\kappa_\alpha$ is a diffusion coefficient that is known to be
comparable to or somewhat below the value of $\etat$ \cite{Mit10,HB10}.

We solve the governing one-dimensional equations \eqs{dmeanBdt}{dalpdt}
using the {\sc Pencil Code}
in cylindrical coordinates, $(r,\phi,z)$, assuming axisymmetry
and homogeneity along the $z$ direction,
$\partial/\partial\phi=\partial/\partial z=0$,
in a one-dimensional domain $0\leq r\leq R$.
On $r=0$ regularity of all functions is obeyed, while on $r=R$
we assume perfect conductor boundary conditions, which implies
that $\nnn\times\meanEE=\nnn\times\meanJJ=\nullvector$, and thus
$\nnn\times\partial\meanAA/\partial t=\nullvector$,
i.e., $\nnn\times\meanAA=\const$.

As initial condition, we choose a uniform magnetic field $B_0$
in the $z$ direction.
In terms of the vector potential, this implies
\EQ
\AAA(r,0)=(0,B_0 r/2,0)
\EN
for the initial value of $\AAA(r,t)$.

We drive the system though the externally applied mean electromotive force
in the $z$ direction.
We choose
\EQ
\meanE^{\rm ext}_z(r)=\meanE^{\rm ext}_0 J_0(k_1r),
\EN
where $\meanE^{\rm ext}_0$ is the value of the mean electromotive force
on the axis and $k_1R\approx2.4048256$ is the rescaled cylindrical wavenumber
for which $\meanE^{\rm ext}_z(R)=0$, which corresponds to the first zero
of the Bessel function of order zero, and thus satisfies the
perfect conductor boundary condition on $r=R$.
An important control parameter of our model is the non-dimensional ratio
\EQ
{\cal Q}=\meanE^{\rm ext}_0/\etat\kf B_0,
\EN
which determines the degree of magnetic helicity injection.
Other control parameters include the normalized strength of the
imposed field,
\EQ
{\cal B}=B_0/\Beq
\EN
and the value of Lundquist number,
\EQ
\LLu=\vA/\eta\kf,
\EN
which is a nondimensional measure of the inverse microscopic magnetic
diffusivity, where $\vA=B_0/\sqrt{\mu_0\rho_0}$ is the Alfv\'en speed
associated with the imposed field.
The Lundquist number also characterizes the ratio of turbulent to
microscopic  magnetic diffusivity, i.e.,
\EQ
\RRey\equiv\etat/\eta=\urms/3\eta\kf=\LLu/3{\cal B},
\EN
which we refer to as the magnetic Reynolds number.
Note that, if we were to define the magnetic Reynolds number as
$\Rm=\urms/\eta\kf$, as is often done, then $\RRey=\Rm/3$ would
be three times smaller.
Finally, the wavenumber of the energy-carrying turbulent eddies is
expressed in terms of the dimensionless value of $\kf R$.
We treat $\kf$ here as an adjustable parameter that characterizes
the degree of scale separation, i.e., the ratio of the scale of the
domain to the characteristic scale of the turbulence.
In most of the cases we consider $\kf R=10$.
In summary, our model is characterized by four parameters:
${\cal Q}$, ${\cal B}$, $\LLu$, and $\kf R$.
In models with magnetic helicity flux we also have the parameter
$\kappa_\alpha/\etat$.

In addition to plotting the resulting profiles of magnetic field and
current density, we also determine mean-field magnetic energy and
helicity, as well as mean-field current helicity, i.e.,
\begin{equation}
M_{\rm m}=\bra{\meanBB^2/2\mu_0},\quad
H_{\rm m}=\bra{\meanAA\cdot\meanBB},\quad
C_{\rm m}=\bra{\meanJJ\cdot\meanBB},
\end{equation}
where $\bra{.}=\int_0^R .\, r\,\dd r/(\half R^2)$ denotes a volume average
and the subscript m refers to mean-field quantities.
Following similar practice of earlier work \cite{BB02,BDS02},
we characterize the solutions further by computing the effective
wavenumber of the mean field, $\kmean$, and the degree $\epsm$ to which
it is helical, via
\EQ
\kmean^2=\mu_0 C_{\rm m}/H_{\rm m},\quad
\epsm=C_{\rm m}/2\kmean M_{\rm m}.
\EN
In the following we shall refer to $\epsm$ as the relative magnetic helicity.
We recall that, even though $\meanAA\cdot\meanBB$ is gauge-dependent,
for perfect conductor boundary conditions, the integral
$\int\meanAA\cdot\meanBB\,\dd V$ is gauge-invariant, and so is then $\kmean$.
Similar definitions also apply to the fluctuating field,
whose current helicity is given by
\begin{equation}
C_{\rm f}=\bra{\alpha}\Beq^2/\mu_0\etat.
\end{equation}
The magnetic helicity of the fluctuating field is then
$H_{\rm f}=\mu_0 C_{\rm f}/\kf^2$.
The magnetic energy of the fluctuating field can be estimated
under the assumption that the field is fully helical, i.e.,
$\bra{\bb^2}=\kf|\bra{\aaaa\cdot\bb}|$, so that
$M_{\rm f}=|C_{\rm f}|/2\kf$.
We study both the steady state case where ${\cal Q}$ and ${\cal B}$
are non-vanishing, and the decaying case where ${\cal Q}={\cal B}=0$.
In the latter case, we monitor the decay rates of the magnetic field.

\section{Results}

\subsection{Driven field-aligned currents}

\begin{figure}[t!]\begin{center}
\includegraphics[width=\columnwidth]{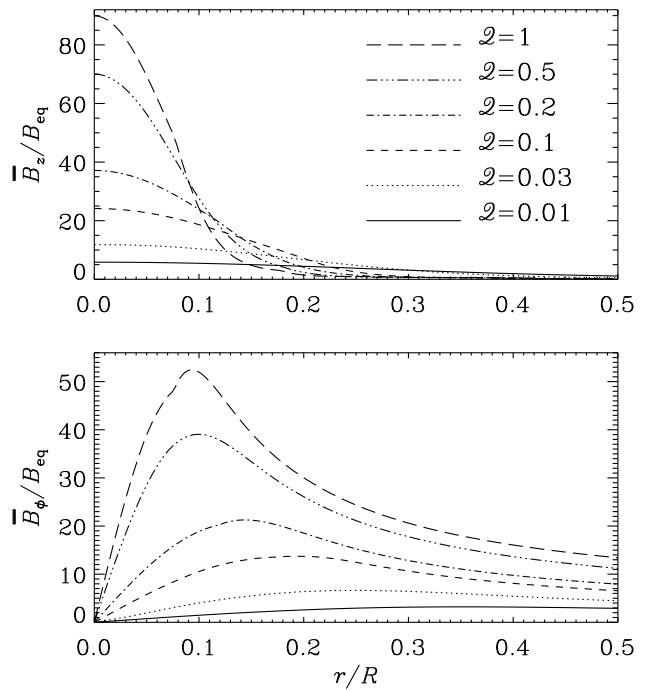}
\end{center}\caption[]{
Equilibrium profiles for three different driving strengths
for ${\cal B}=1$, $\LLu=1000$, and $\kf R=10$.
}\label{pbzbp_comp}\end{figure}

\begin{figure}[t!]\begin{center}
\includegraphics[width=\columnwidth]{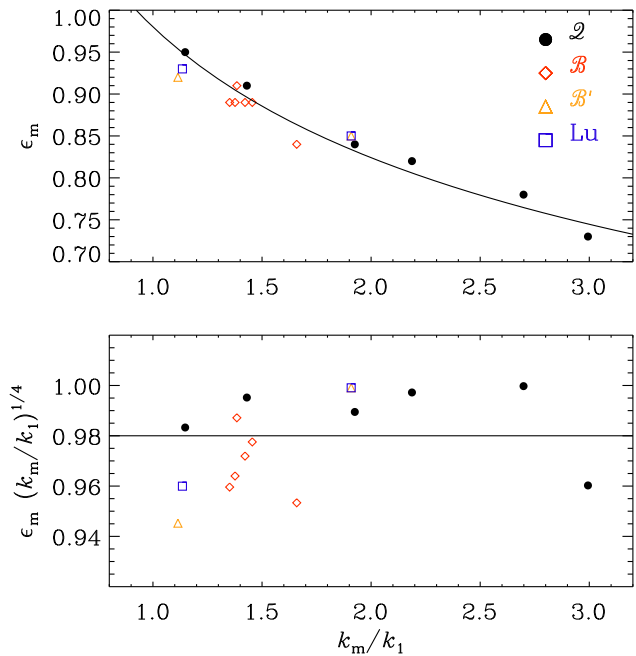}
\end{center}\caption[]{
(Color online) Dependence of $\epsm$ on $\kmean$ for different sets
of runs where either ${\cal Q}$ is varied (filled symbols),
${\cal B}$ is varied while keeping $\RRey=100$ (red diamonds),
${\cal B}$ is varied while keeping $\LLu=100$ (orange triangles),
or $\LLu$ is varied (blue squares).
}\label{pkeps}\end{figure}

We begin by considering the case without magnetic helicity fluxes and
take ${\cal B}=1$, $\LLu=1000$ (corresponding to $\RRey=333$) and $\kf R=10$.
The resulting values of $\kmean$ and $\epsm$ are given in \Tab{TQ}
and the mean magnetic field profiles are compared in \Fig{pbzbp_comp}
for different values of ${\cal Q}$.
It turns out that, as we increase the value of ${\cal Q}$, the magnetic
helicity of the mean field increases, i.e.\ the product $\epsm\kmean$
increases, but the {\it relative} helicity of the mean magnetic field
decreases slightly, i.e., $\epsm$ decreases.
The value of $\kmean$ increases with ${\cal Q}$, which means that the mean
field will be confined to a progressively thinner core around the axis.
Furthermore, the anti-correlation between $\epsm$ and $\kmean$ is also found
when varying ${\cal B}$ (see \Tab{TB}), $\LLu$, or $\RRey$.
This is demonstrated in \Fig{pkeps}, where we show that
$\epsm$ is in fact proportional to $(\kmean/k_1)^{-1/4}$ and that the
product $\epsm(\kmean/k_1)^{1/4}$ is approximately constant,
even though ${\cal Q}$, ${\cal B}$, or $\LLu$ are varied.
This scaling is unexpected and there is currently no theoretical
interpretation for this behavior.

\begin{table}[b!]\caption{
${\cal Q}$ dependence of $\kmean$ and $\epsm$
for ${\cal B}=1$, $\LLu=1000$, and $\kf R=10$.
}\vspace{12pt}\centerline{\begin{tabular}{lccccccc}
${\cal Q}\;$&0.01\ &\ 0.03\ &\ 0.10\ &\ 0.20\ &\ 0.50\ &\ 1.00 \\ \hline
$\kmean R\;$ &2.76 \ &\ 3.44\ &\ 4.63\ &\ 5.26\ &\ 6.49\ &\ 7.20 \\
$\epsm\;$  &0.95 \ &\ 0.91\ &\ 0.84\ &\ 0.82\ &\ 0.78\ &\ 0.73
\label{TQ}\end{tabular}}\end{table}

It is interesting to note that $\kmean$ does not vary significantly
with ${\cal B}$, provided $\RRey$ is held fixed.
However, for weak fields, e.g., for ${\cal B}=0.1$, the dynamics of the mean
field is no longer controlled by magnetic helicity evolution, and the value of
$\kmean R$ has then dropped suddenly by nearly a factor of 2, and $\epsm$
is in that case no longer anti-correlated with $\kmean$.
This data point falls outside the plot range of \Fig{pkeps}, 
and is therefore not included.
Also, if only $\LLu$ is held fixed, so that $\RRey$ varies
with ${\cal B}$, then $\kmean$ is no longer weakly varying with
${\cal B}$, and varies more strongly in that case.

\begin{table}[b!]\caption{
${\cal B}$ dependence of $\kmean$ and $\epsm$
for ${\cal Q}=0.1$, $\RRey=100$, and $\kf R=10$.
}\vspace{12pt}\centerline{\begin{tabular}{lccccccc}
${\cal B}\;$&0.1 \ &\ 0.2 \ &\ 0.5 \ &\  1  \ &\  2  \ &\  5   &  10 \\ \hline
$\kmean R\;$&1.80 \ &\ 3.33\ &\ 3.50\ &\ 3.99\ &\ 3.42\ &\ 3.31 & 3.25\\
$\epsm\;$  &0.51 \ &\ 0.91\ &\ 0.89\ &\ 0.84\ &\ 0.89\ &\ 0.89 & 0.89\\
\label{TB}\end{tabular}}\end{table}

We must ask ourselves why the axial field component does not show a
reversal in radius, as is the case in the RFP.
Experimental studies of the RFP provide direct evidence for a reversal.
By comparing radial profiles of the axial current,
$\meanJ_\parallel/\sigma$, with those of the axial electric
field, $\meanE_\parallel$, one concludes
that the mismatch between the two must come from the
$\meanemf_\parallel$ term \cite{JP02,HPS89}.
These studies show that $\meanE_\parallel<\meanJ_\parallel/\sigma$
near the axis and $\meanE_\parallel>\meanJ_\parallel/\sigma$ away
from it (assuming $\meanB_\parallel>0$ on the axis).
Comparing with \Eq{OhmMean}, it is therefore clear that $\meanemf_\parallel$
must then be negative near the axis and positive near the outer rim.
Turning now to dynamo theory, it should be emphasized
that there are two contributions to $\meanemf_\parallel$,
one from $\alpha\meanBB$ and one from $-\etat\mu_0\meanJJ$; see \Eq{meanEMF}.
Let us therefore discuss in the following the expected sign of
$\meanemf_\parallel$.
Given that ${\cal Q}$ is positive, $\meanJJ\cdot\meanBB$ must also be
positive, and therefore we expect $\alpha$ to be positive.
If the mean magnetic field was really sustained by a dynamo,
the $\alpha$ term would dominate over the $\etat$ term, but this
is likely not the case here.
Indeed, by manipulating \Eq{dalpdt} we see that, in the steady state
without magnetic helicity fluxes, the equation for $\alpha$ takes the form
\EQ
\alpha={\RRey\etat\mu_0\meanJJ\cdot\meanBB/\Beq^2\over1+\RRey\meanBB^2/\Beq^2};
\label{alpha_steady}
\EN
see, e.g., Ref.~\cite{BB02}.
However, as alluded to above, the relevant term entering $\meanEMF$ is the
combination $\alpha_{\rm red}=\alpha-\etat\mu_0\meanJJ\cdot\meanBB/\meanBB^2$,
which is the reduced $\alpha$.
Inserting \Eq{alpha_steady} yields
\EQ
\alpha_{\rm red}=
-{\etat\mu_0\meanJJ\cdot\meanBB/\meanBB^2\over1+\RRey\meanBB^2/\Beq^2},
\label{alpha_reduced}
\EN
with a minus sign in front.
The important point here is that $\alpha_{\rm red}$ is indeed negative
if $\meanJJ\cdot\meanBB$ is positive.
This means that we can only expect $\meanE_\parallel<\meanJ_\parallel/\sigma$,
which is the situation in the RFP near the axis \cite{HPS89}.
In order to reverse the ordering and to produce a reversal of the
axial field, one would need to have an $\alpha$ effect that dominates
over turbulent diffusion.
Note also that for strong mean fields, $\alpha_{\rm red}$ is of the order
of the microscopic magnetic diffusivity.
(This situation is well-known for nonlinear dynamos, because there
$\alpha_{\rm red}$ and the microscopic diffusion term $\eta\kmean$
have to balance each other in a steady state \cite{BRRS08}.)

We note in passing that $\kmean$ enters neither in \Eq{alpha_steady}
nor in \Eq{alpha_reduced}.
However, $\kmean$ does enter if there is a magnetic helicity flux and
it affects the time-dependent case, as is also clear from \Eq{dalpdt}.
Both cases will be considered below.

\subsection{Effect of magnetic helicity flux}

Next, we study cases where a diffusive magnetic helicity flux
is included.
In our model with perfectly conducting boundaries, the magnetic helicity
flux vanishes on the boundaries, so no magnetic helicity is exported
from the domain, but the divergence of the flux is finite and can thus
modify the magnetic $\alpha$ effect.
The same is true of periodic boundaries, where no magnetic helicity is
exported, but the flux divergence is finite and can alleviate catastrophic
quenching in dynamos driven by the kinetic $\alpha$ effect \cite{HB11}.

\begin{figure}[t!]\begin{center}
\includegraphics[width=\columnwidth]{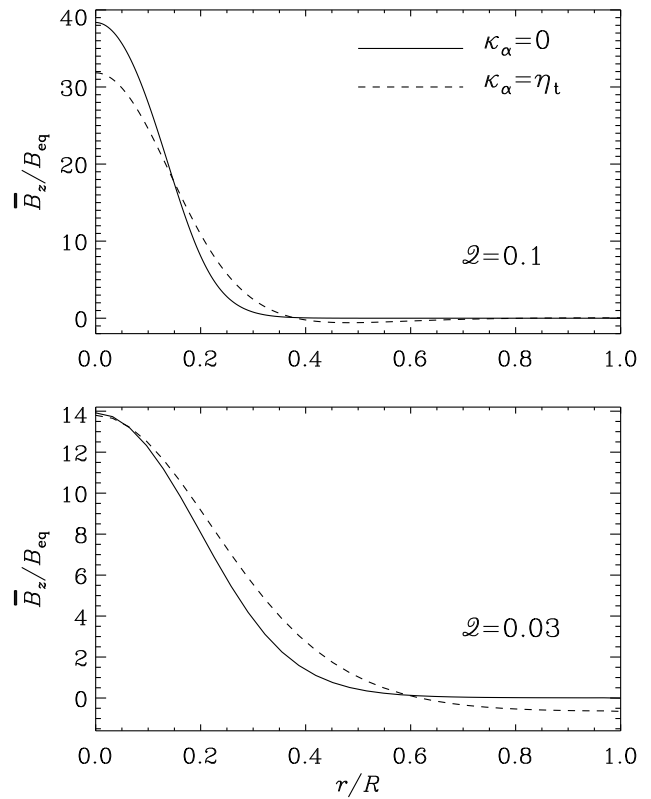}
\end{center}\caption[]{
Effect of magnetic helicity flux on equilibrium profiles of $\meanB_z$
for ${\cal Q}=0.1$ (upper panel) and ${\cal Q}=0.03$ (lower panel).
Note the field reversal at the outer rim in the latter case.
}\label{pflux}\end{figure}

In \Fig{pflux} we compare profiles of $\meanB_z$ with and without
magnetic helicity flux.
It turns out that the $\kappa_\alpha$ term has the effect of smoothing out
the resulting profile of $\meanB_z$.
More interestingly, it can lead to a reversal of $\meanB_z$ at
intermediate radii.
For our reference run with ${\cal Q}=0.1$ (upper panel),
the reversal is virtually absent at the rim of the cylinder.
This is mainly because the pinch is so narrow; see \Tab{TF}.
However, when decreasing ${\cal Q}$ to 0.03, there is a clear
reversal also at the outer rim (lower panel).
However, decreasing ${\cal Q}$ to 0.01 does not increase the
extent of the reversal.
In none of these cases the field reversal is connected with a change
of sign of $\alpha_{\rm red}$.
Instead, $\alpha_{\rm red}$ is always found to be negative, even in
the presence of a magnetic helicity flux.
Thus, the sign reversal of $\meanB_z$ is therefore associated with
a sign reversal of $\meanJ_z$ at the same radius.
Nevertheless, the reversal is still not very strong with
$\min(\meanB_z)/\max(\meanB_z)\approx-0.07$, while in laboratory RFPs
this ratio is typically $-0.2$ \cite{HPS89}.

\begin{table}[b!]\caption{
Values of $\kmean$ and $\epsm$ with and without magnetic helicity flux.
}\vspace{12pt}\centerline{\begin{tabular}{l|cc|ccccc}
${\cal Q}\;$ & \multicolumn{2}{c|}{0.03} & \multicolumn{2}{c}{0.1} \\
\hline
$\kappa_\alpha/\etat$ & 0 & 1 & 0 & 1 \\
\hline
$\kmean R\;$& 4.63 & 4.50 & 3.51 & 3.32 \\
$\epsm\;$   & 0.84 & 0.83 & 0.91 & 0.92 \\
\label{TF}\end{tabular}}\end{table}

\subsection{Decay calculations}

\begin{figure}[t!]\begin{center}
\includegraphics[width=\columnwidth]{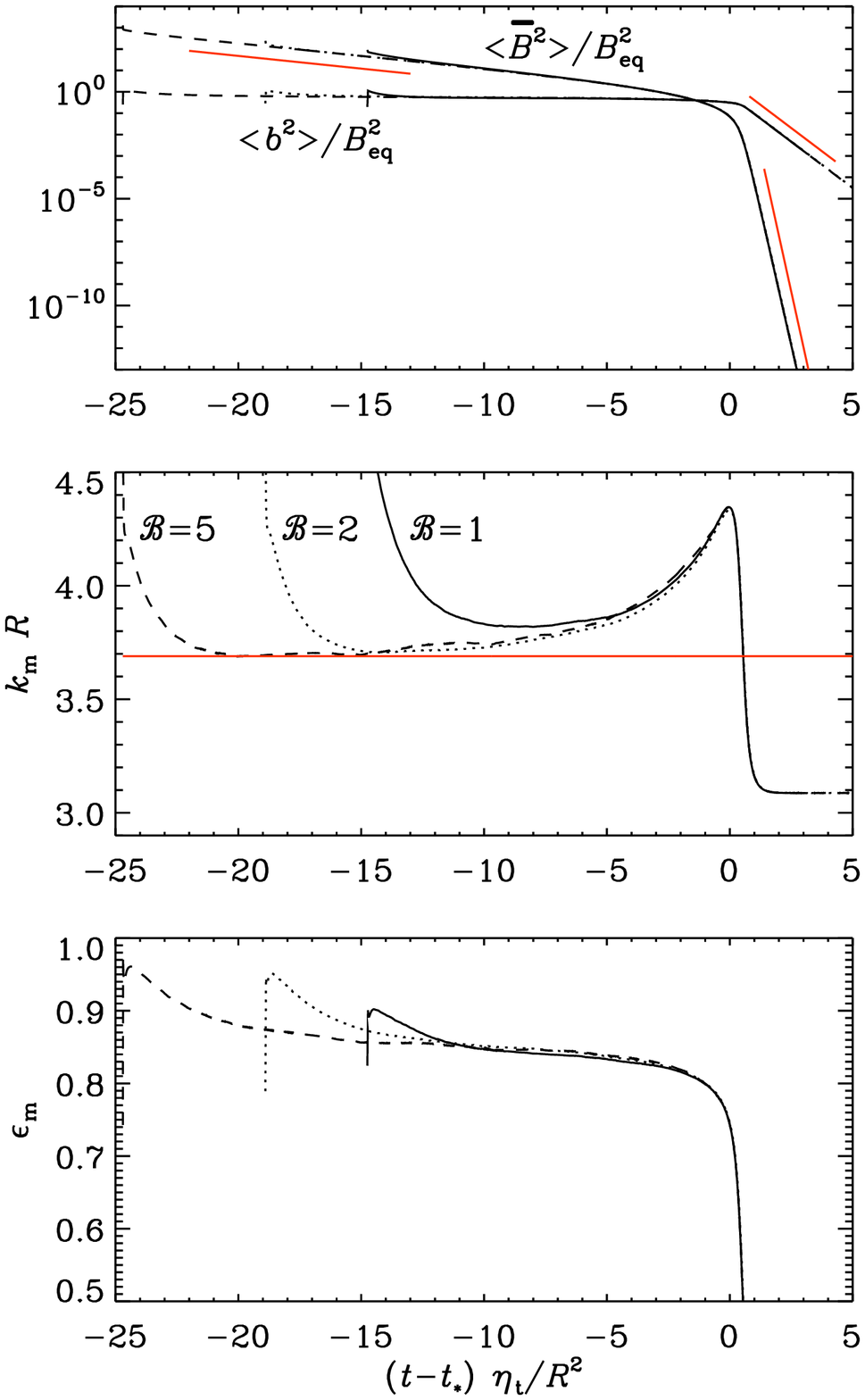}
\end{center}\caption[]{
(Color online)
Evolution of $\bra{\meanBB^2}/\Beq^2$, $\kmean/k_1$, and $\epsm$
for different values of ${\cal B}$.
Note that time is shifted by $t_*$, which is the time when
$\kmean$ attains its second maximum.
In the top panel, the red lines indicate
resistive decay rate of the large-scale field at early times,
resistive decay rate of the small-scale field at late times, and
the turbulent decay rate of the large-scale field at late times.
}\label{pdecay}\end{figure}

Next, we consider the case of a decaying magnetic field in the
absence of an external electric field.
In that case all components of $\meanBB$ must eventually decay to zero.
The evolution of the magnetic energy of the resulting mean and
fluctuating fields is shown in \Fig{pdecay}, together with the
evolution of $\kmean$ and $\epsm$.
At early times, when $\bra{\meanBB^2}\gg\Beq^2$, the energy of the
large-scale magnetic field decays at the resistive rate
$\lambda=-2\eta\kmean^2$.
During that time, the energy of the small-scale field stays
approximately constant: the magnetically generated $\alpha$ effect
almost exactly balances turbulent diffusion and the magnetic field
can only decay at the resistive rate.
However, at later times, when $\bra{\meanBB^2}\ll\Beq^2$, the energy of the
small-scale field decays with a negative growth rate $\lambda=-2\eta\kf^2$,
which then speeds up the decay of the energy of the large-scale magnetic field
to a rate that is about $1.3\times\etat\kmean^2$, where we have used
the value $\kmean R=3.1$ that is relevant for the late-time decay.
This value is also that obeying Taylor's \cite{Tay74} postulated
minimum energy state.
Again, no reversal of the magnetic field is found, except in cases
where there is an internal magnetic helicity flux in the system.

\section{Conclusions}

The present work is an application of the dynamical quenching
model of modern mean-field dynamo theory to magnetically driven and
decaying turbulence in cylindrical geometry.
In the driven case, an external electric field is applied,
which leads to magnetic helicity injection at large scales.
Such a situation has not yet been considered in the framework
of mean-field theory.
It turns out that in such a case there is a weak anti-correlation
between the actual value of magnetic helicity of the mean field,
$\epsm\kmean$, and the relative magnetic helicity $\epsm$ with
$\epsm\sim\kmean^{-1/4}$.
This weak anti-correlation is found to be independent of whether
${\cal Q}$, ${\cal B}$, or $\LLu$ are varied.
No theoretical interpretation of this behavior has yet been offered.
In the decaying case, we find that the decay rate is close to the
resistive value when the field is strong, i.e., $\meanB>\Beq$, and drops
to the turbulent resistive value when the mean field becomes weaker.
This behavior has also been found in earlier calculation of the decay
of helical magnetic fields in Cartesian geometry \cite{YBR03}.

The original anticipation was that our model reproduces some features
of the RFP that is studied in connection with
fusion plasma confinement.
It turns out that the expected field reversal that gives the RFP its name
is found to require the presence of magnetic helicity fluxes.
Without such fluxes, there is no reversal; see \Fig{pflux}.
Nevertheless, the reversal is rather weak compared with laboratory RFPs.
This discrepancy can have several reasons.
On the one hand, we have been working here with a model that has
previously only been tested under simplifying circumstances in which
there is turbulent dynamo action driven by kinetic helicity supply.
It is therefore possible that the model has shortcomings that
have not yet been fully understood.
A related possibility is that the model is still basically valid,
but our application to the RFP has been too crude.
For example, the assumption of fixed values of $\etat$ and $\Beq$
is certainly quite simplistic.
On the other hand, it is not clear that this simplification would really
affect the outcome of the model in any decisive way.
A different possibility is that the application of an external
electric field is not representative of the RFP.
However, an important basic idea behind the present setup has been to
establish a model as simple as possible, that could be tested by performing
corresponding three-dimensional simulations of a similar setup.
This has not yet been attempted, but this would clearly constitute
a natural next step to take.

\acknowledgements

This work was supported in part by the Swedish Research Council,
grant 621-2007-4064, the European Research Council under the
AstroDyn Research Project 227952,
and the National Science Foundation under Grant No.\ NSF PHY05-51164.
HJ acknowledges support from the U.S.\ Department of Energy's Office of
Science -- Fusion Energy Sciences Program.


\begin{thebibliography}{}

\bibitem{Mof78}
H. K. Moffatt\ybook{1978}
{Magnetic field generation in electrically conducting fluids}
{Cambridge University Press, Cambridge}

\bibitem{KR80}
F. Krause and K.-H. R\"adler\ybook{1980}
{Mean-field magneto\-hydro\-dy\-na\-mics and dynamo theory}
{Pergamon Press, Oxford}

\bibitem{Ogino}
T. Ogino\yjgr{1986}{91}{6791}

\bibitem{Ji_etal96}
H. Ji, S. C. Prager, A. F. Almagri, J. S. Sarff, and H. Toyama\ypp{1996}{3}{1935}

\bibitem{Cadarache}
R. Monchaux, M. Berhanu, M. Bourgoin, et al.\yprl{2007}{98}{044502}

\bibitem{BN80}
H. A. B. Bodin and A. A. Newton\yjour{1980}{Nuclear Fusion}{20}{1255}

\bibitem{Tay86}
J. B. Taylor\yprl{1986}{58}{741}

\bibitem{BJ06}
E. G. Blackman and H. Ji\ymn{2006}{369}{1837}

\bibitem{PFL76}
A. Pouquet, U. Frisch, and J. L\'eorat\yjfm{1976}{77}{321}

\bibitem{FB02}
G. B. Field and E. G. Blackman\yapj{2002}{572}{685}

\bibitem{BB02}
E. G. Blackman and A. Brandenburg\yapj{2002}{579}{359}

\bibitem{Sub02}
K. Subramanian\yjour{2002}{Bull.\ Astr.\ Soc.\ India}{30}{715}

\bibitem{KR82}
N. I. Kleeorin and A. A. Ruzmaikin\yjour{1982}{Magne\-to\-hydro\-dynamics}{18}{116}

\bibitem{BS05}
A. Brandenburg and K. Subramanian\yan{2005}{326}{400}

\bibitem{BS04}
A. Brandenburg and C. Sandin\yana{2004}{427}{13}

\bibitem{B05}
A. Brandenburg\yapj{2005}{625}{539}

\bibitem{JP02}
H. Ji and S. C. Prager\yjour{2002}{Magnetohydrodynamics}{38}{191}
{\sf astro-ph/0110352}

\bibitem{Tay74}
J. B. Taylor\yprl{1974}{33}{1139}

\bibitem{AB84}
A. Y. Aydemir and D. C. Barnes\yprl{1984}{52}{930}

\bibitem{YBR03}
T. A. Yousef, A. Brandenburg, and G. R\"udiger\yana{2003}{411}{321}

\bibitem{SCN85}
D. D. Schnack, E. J. Caramana, and R. A. Nebel\ypf{1985}{28}{321}

\bibitem{SMN96}
H.-E. S\"atherblom, S. Mazur, and P. Nordlund\yppcf{1996}{38}{2205}

\bibitem{CBE06}
S. Cappello, D. Bonfiglio, and D. F. Escande\ypp{2006}{13}{056102}

\bibitem{Str85}
H. R. Strauss\ypf{1985}{28}{2786}

\bibitem{BH86}
A. Bhattacharjee and E. Hameiri\yprl{1986}{57}{206}

\bibitem{Ji99}
H. Ji\yprl{1999}{83}{3198}

\bibitem{KRR95}
N. Kleeorin, I. Rogachevskii, and A. Ruzmaikin\yana{1995}{297}{159}

\bibitem{SB06}
K. Subramanian and A. Brandenburg\yapj{2006}{648}{L71}

\bibitem{HB10}
A. Hubbard and A. Brandenburg\ygafd{2010}{104}{577}

\bibitem{Mit10}
D. Mitra, S. Candelaresi, P. Chatterjee, R. Tavakol, and A. Brandenburg\yan{2010}{331}{130}

\bibitem{BDS02}
A. Brandenburg, W. Dobler, and K. Subramanian\yan{2002}{323}{99}

\bibitem{HPS89}
Y. L. Ho, S. C. Prager,  D. D.Schnack\yprl{1989}{62}{1504}

\bibitem{BRRS08}
A. Brandenburg, K.-H. R\"adler, M. Rheinhardt, and K. Subramanian\yapj{2008}{676}{740}

\bibitem{HB11}
A. Hubbard and A. Brandenburg\yapj{2011}{727}{11}

\end{thebibliography}
\end{document}